# Non-reciprocal absorption and zero reflection in physically separated dual photonic resonators by traveling-wave-induced indirect coupling

Bojong Kim, Junyoung Kim, Hae-Chan Jeon, and Sang-Koog Kim

*National Creative Research Initiative Center for Spin Dynamics and Spin-Wave Devices, Nanospinics Laboratory, Research Institute of Advanced Materials, Department of Materials Science and Engineering, Seoul National University, Seoul 151-744, Republic of Korea*

We experimentally explored novel behaviors of non-reciprocal absorption and almost zero reflection in a dual photon resonator system, which is physically separated and composed of two inverted split ring resonators (ISRRs) with varying inter-distances.  We also found that an electromagnetically-induced-transparency (EIT)-like peak at a specific inter-distance of $d =$ 18 mm through traveling waves flowing along a shared microstrip line to which the dual ISRRs are dissipatively coupled. With the aid of CST-simulations and analytical modeling, we found that destructive and/or constructive interferences in traveling waves, indirectly coupled to each ISRR, result in a traveling-wave-induced transparency peak within a narrow window. Furthermore, we observed not only strong non-reciprocal responses of reflectivity and absorptivity at individual inter-distances exactly at the corresponding EIT-like peak positions, but also nearly zero reflection and almost perfect absorption for a specific case of $d =$ 20 mm. Finally,  the unidirectional absorptions with zero reflection at $d =$ 20 mm are found to be ascribed to a non-Hermitian origin.  This work not only provides a better understanding of traveling-wave-induced indirect coupling between two photonic resonators without magnetic coupling, but also suggests potential implications for the resulting non-reciprocal behaviors of absorption and reflection in microwave circuits and quantum information devices.

A) Correspondence and requests for materials should be addressed to S.-K. K (sangkoog@snu.ac.kr).

## I. INTRODUCTION

The propagation-direction dependence of wave interferences has emerged as a crucial research topic due to its broad applicability across electrodynamics, acoustics, matter waves, and quantum electronic [1–4]. The study of transmitted photons and their role in signal transport, filtering, and light-matter interaction has attracted significant interest due to their potential in quantum information processing technologies. Attentions has been focused on manipulating transmitted photons in related fields like metamaterials, where significant progress has been made in achieving perfect absorption and transmission, and electromagnetically induced transparency and absorption (EIT/EIA) [5–8]. However, reflected photons can be detrimental, not only reducing the efficiency of transmission and absorption but also interfering with excitation and adversely affecting the performance of entire optical systems, e.g., the frequency stability of an excitation laser [9]. Consequently, the development of structures that minimize reflection and tailor transmission and absorption is essential for specific applications.

In order to achieve zero reflection, transformation optics can provide a promising solution using inhomogeneous, anisotropic materials (i.e., transformation media). Within the context of photonics, such significant advancements have the potential to catalyze the development of a new generation of optical isolators, circulators, unidirectional biosensors, switches, and modulators [1,2,4]. Additionally, optical systems with parity-time (PT) symmetry offer another promising approach to suppress reflection [10,11]. Therefore, PT symmetry has garnered significant attention in the field of metamaterials, as it enables non-Hermitian Hamiltonians to exhibit real energy spectra, thereby enabling classical systems to exhibit quantum mechanical properties. Due to the inherent equivalence between the Schrödinger equation in quantum mechanics and the wave equation in optics, the investigation of combined PT symmetry has received extensive focus within classical optical systems exhibiting non-Hermitian optical potentials. In a non-Hermitian Hamiltonian governing an open quantum system, the exceptional point (EP) – the merging of two eigenvalues and their corresponding eigenvectors, represents one of the most important degeneracies. So far, PT optical materials have exhibited various unique phenomena, such as single-mode lasers [12], coherent prefect absorbers [6,13], optical non-reciprocal propagation [14–16], cloaking objects [17,18], unidirectional reflectionlessness [10,11,19–25], and more.

Recently, considerable attention has been directed towards investigating unidirectional reflectionless phenomena based on balanced gain and loss [10,18,21,22]. Furthermore, unidirectional reflectionlessness has been investigated in non-Hermitian systems without balanced gain and loss [19,24]. Lin et al. [10] theoretically demonstrated unidirectional invisibility at the EP in PT-symmetric periodic structures. Feng et al. [11] demonstrated experimentally the occurrence of unidirectional light reflection at the exceptional point (EP) within a passive waveguide structure. These findings suggest that unidirectional light reflection stems from EPs present in both passive and PT-symmetric systems. On the other hand, there has been a shift in focus towards exploring the interaction between resonators without considering PT-symmetry. This approach also enables unidirectional reflectionless absorption and transmission without requiring complicated structural designs.

In this study, to achieve a well-defined indirect coupling between separated resonators through a single-channel waveguide, similar to dielectric waveguide optics[41], we fabricated a compact microwave device. This device comprises two inverted split ring resonators (ISRRs) physically separated at sufficiently distinct inter-distances, but indirectly coupled via a common microstrip line (Fig. 1). We achieved unidirectional (non-reciprocal) zero reflection and nearly perfect absorption from the simple dual ISRRs at certain specific inter-distances.

The dual ISRRs can exchange energies with the shared environment and indirectly couple to each other via the travelling waves flowing along the microstrip line. By adjusting the inter-distance of the dual ISRRs, we can simply manipulate the constructive and destructive interference of the traveling waves to achieve traveling wave induced transparency similar to EIT. Also, the experimentally observed nearly zero reflection with high non-reciprocity is ascribed to the destructive interference between the direct reflection by the first ISRR and the multiple reflection involving the second ISRR. Strictly destructive interference occurs only at a single frequency , i.e., traveling-wave-induced transparency peak, for a given specific inter-distance. Our work not only provides a better understanding of the traveling-wave-induced indirect coupling effect leading to novel unidirectional, reflectionless absorption at the exactly corresponding EIT-like peak depending on the inter-distance, but also has potential implications in microwave circuits and quantum information devices.

## II. Sample's geometry and measurements of scattering parameters

Fig. 1 illustrates the dimensions and geometry of samples consisting of two ISRRs with different split-gap orientations relative to a given microstrip-line axis. The details of the dimensions are given in caption of Fig.1. The dimensions of each ISRR were the same as those reported in [26]. Since the two ISRRs of the same gap orientations (i.e., both parallel or both perpendicular to the microstrip line) exhibit an identical single resonant frequency, we intentionally positioned the split gaps of the two ISRRs differently [27,28]. The different split gap orientations (here 90 versus 0 degree) with respect to the microstrip line axis result in different resonant frequencies, $\omega_1/2\pi$ = 3.8 GHz and $\omega_2/2\pi$ = 4.2 GHz according to CST-simulation. Additionally, we ensured that the separation distances between the dual ISRRs were sufficiently large to prevent direct coupling between the two ISRRs. We conformed that the splitting of the dual ISRRs' modes vanishes when $d > 8$ mm, indicating that the direct coupling between the dual ISRRs is prevented [29]. Therefore, we chose specific distances of $d =$ 18,19, and 20 mm for the dual-ISRR system.

We experimentally measured the transmission ($S_{21}$ and $S_{12}$) and reflection ($S_{11}$ and $S_{22}$) scattering parameters from both directions, port 1 to port 2 and vice versa, by applying oscillating AC currents of microwave frequency ($f_{AC} = \omega/2\pi$) to the microstrip feeding line using a vector network analyzer (VNA, Agilent PNA series E8362C). For more details on the sample preparation and measurements, see our earlier publications [26,29].

## III. RESULTS AND DISCUSSION
### A. Observation of EIT-like peaks

Figure 2(a) displays the transmission $|S_{21}|$ and $|S_{12}|$ spectra versus AC current frequency from port 1 to port 2 and vice versa for $d =$ 18, 19, and 20 mm. The direction-dependent transmission spectra show nearly identical spectra for each $d$ value, indicating these samples exhibit reciprocal transmission characteristics for each $d$ value, with no shift in the lower and higher-branch frequencies (see two dips). However, the transmission powers for the lower and higher frequencies exhibit different asymmetries according to $d$. This asymmetry increases from $d =$ 18 mm through $d =$ 19 mm to $d =$ 20 mm, even with a 1mm change in the inter-distance. Most interesting is the transparency peak (marked by green-color circle) between the lower and higher-absorption peaks; the transparency peak becomes high when $d$ approaches 18 mm. This type of transparency is referred to as an analogue of EIT observed in metamaterials. The EIT in waveguide-resonator systems is manifested through the coupling of a radiative (bright) resonator to a sub-radiant (dark) resonator [8,30–32], originating from the near-field couplings between two metamaterial-based resonators. The radiative resonator actively engages with the propagating modes of a given waveguide, while the sub-radiant resonator remains uncoupled, exerting minimal influence on light transmission and reflection. However, in our dual-ISRR system, since the inter-distances ($d$ = 18, 19, and 20 mm) of the dual resonators are sufficiently large to prevent direct coupling, the observed EIT-like peak is rather due to indirect coupling caused by traveling waves in the microstrip line, with which the dual resonators interact dissipatively.

To further understand the entire spectra of $S_{21}$ and $S_{12}$ quantitatively, we conducted simulations using CST Microwave Studio and an analytical modeling using a coupled mode theory (details of the analytical model will be explained in a later session). Figure 2(b) illustrates the resulting transmission spectra for specific inter-distances, $d$ =18,19, and 20; simulation (black circle) and analytical modeling (blue circle). These results are in good agreement, also reproducing the experimental spectra shown in Fig. 2(a) well, in terms of the power and shape of the transmission spectra, except for a slight difference in the corresponding resonance frequencies. This discrepancy in the resonance peaks can be ascribed to the impedance mismatches resulting from the coaxial-to-microstrip line transitions at both connectors' ends and/or mismatches in dimensions between the real samples and the model system. For a finer step of $d$ = 0.2 mm between $d$ = 18 and 20 mm, we also contour-plotted the simulation and analytical modeling, as shown in Fig. 2(c). The transparency peak line (in red) is clearly present between the two absorption lines (in blue), and it shifts towards higher frequencies as the inter-distance decreases. The observed transparency peaks are similar to EIT peaks seen in bright-dark resonator systems due to near-field coupling, but in our case, they are attributed to indirect coupling through traveling waves. According to the Fabry–Pérot theoretical model, the signal transmission is enhanced when the Fabry–Pérot resonant condition is satisfied [33–36].

To further examine the reflection spectra related to the transparency peaks, we experimentally measured the reflection $|S_{11}|$ and $|S_{22}|$ spectra in both directions, from port 1 to port 2 and vice versa, as shown in Fig. 3(a). Unlike the transmission spectra, the reflection spectra exhibit non-reciprocal behavior between both directions. This non-reciprocal reflection reverses for $d = 18$ and 19 mm, even with a variation of $d = 1$ mm. Specifically, for $d = 18$ mm the dip in $|S_{11}|$ is greater than that in $|S_{22}|$. However, the dip in $|S_{11}|$ becomes smaller for $d = 19$ mm, and then finally, the dip in $|S_{11}|$ disappears at $d = 20$ mm, indicating nearly perfect reflection for $S_{11}$. As for $|S_{22}|$, the dip is at its maximum for $d = 19$ mm. The direction-dependent nature of the reflection and its significant variation with the inter-distance between the dual ISRRs provide strong evidence of non-reciprocal reflection behavior. These experimental findings strongly suggest that in the design of the two resonators, the inter-distance does not only exhibit a high degree of sensitivity to transmission and reflection characteristics; it also highlights the pivotal role played by indirect coupling, due to the traveling waves flowing along a common microstrip line.

In order to further elucidate the non-reciprocal behavior of reflection, we calculated reflectivity, absorptivity, and transmittivity, according to $T_{21} = |S_{21}|^2$, $T_{12} = |S_{12}|^2$, $R_{11} = |S_{11}|^2$ and $R_{22} = |S_{22}|^2$ and $A_{21} = 1 - R_{11} - T_{21}$ and $A_{12} = 1 - R_{22} - T_{12}$ from the experimentally measured data of transmission $|S_{21}|$, $|S_{12}|$ and reflection $|S_{11}|$, $|S_{22}|$ spectra. In Fig.3(b), the resultant reflectivity (black) and absorptivity (red) spectra show strong, contrasting non-reciprocity even with a variation of $d = 1$ mm for $d = 18$, 19, and 20 mm. Meanwhile, the transmittivity does not show nonreciprocity since $T_{21} \approx T_{12}$, but its magnitude decreases with $d$, as 0.18, 0.12, and 0.035 at $d = 18, 19$, and 20 mm, respectively. In contrast to transmittivity, in the forward direction both the absorptivity ($A_{21}$) and reflectivity ($R_{11}$) reach their maximum values and minimum values respectively, for $d = 18$ mm, and then decrease continuously with $d$. In the backward direction (bottom row in Fig. 3(b)), $A_{12}$ and $R_{22}$ have maximum peaks at $d$ = 19 mm. In particular, at $d$ = 20 mm, we observed weak and strong peaks for $A_{12}$ and $R_{22}$, respectively. However, we found nearly unity(close to one) for $A_{21}$, indicating a strong signal, and no peaks were observed for $R_{11}$. This unidirectional, reflectionless behavior is of great interest in terms of optical, microwave invisibility. Therefore, we will discuss it in more depth using analytical calculations of the scattering matrix combined with a comprehensive analysis of the observed unidirectional absorption and zero reflection.

### B. Analytical model

To quantitatively analyze non-reciprocal reflection and absorption behaviors including EIT-like peak for different $d$ values, we developed an analytical model based on a temporal coupled mode theory [37,38]. This model consists of symmetrical, aperture-side-coupled, dual photon resonators with a single shared transmission line, as schematically illustrated in Fig. 4. When microwave AC currents are injected into the microstrip feeding line from port 1 to port 2 or vice versa, both dynamic electric and magnetic field components are generated by the flowing ac currents. Here, we used two independent ISRRs with different split-gap orientations with respect to a given microstrip-line axis. In the case where the split gap orientation is perpendicular to the microstrip line, the electric field dominates the photon-mode excitation among the electric and magnetic fields generated from the microstrip line. However, when the split gap orientation is parallel to the microstrip line, both electric and magnetic fields participate in the photon-mode excitation, referred to as a cross-polarization effect [28]. The split-gap position dependence strongly modifies the photon-photon coupling, as reported in Ref [29]. In this model, the dual ISRRs separated by an inter-distance ($d$) is considered as two-coupled LC resonators. With the harmonic time dependence of $e^{-j\omega t}$, the temporal normalized mode amplitudes $a_i$ of the ith resonator (i = 1, 2) can be described as follows:

$$\frac{da_i}{dt} = (-j\omega_i - \gamma_i - \kappa_i)a_i + e^{j\theta_i}\sqrt{\kappa_i}S_{f,in}^{(i)} + e^{j\theta_i}\sqrt{\kappa_i}S_{b,in}^{(i)} \tag{1}$$

where $\omega_i$ represents the resonance frequency, $\gamma_i$ is the intrinsic decay rate due to internal loss, and $\kappa_i$ is the extrinsic decay rate due to the coupling with the microstrip line. The decay rates satisfy such relationships as $\gamma_i = \omega_i/(2Q_\gamma)$ and $\kappa_i = \omega_i/(2Q_\kappa)$, where $Q_\gamma$ and $Q_\kappa$ stand for the intrinsic and coupling induced (extrinsic) quality factor of each resonator, respectively. $\theta_i$ is the phase of ISRR-ISRR coupling coefficient. The amplitudes of the incoming and outgoing waves in the common microstrip line are depicted as $S_{f(b),in}^{(i)}$, and $S_{f(b),out}^{(i)}$ (i = 1,2). The subscript f(b) represents the forward (backward) direction of propagating wave modes, as shown in Fig. 4. Through the energy conservation, the outgoing waves of the ith resonator can be written as

$$S_{f,out}^{(i)} = S_{f,in}^{(i)} - e^{j\theta_i}\sqrt{\kappa_i}a_i \tag{2a}$$

$$S_{b,out}^{(i)} = S_{b,in}^{(i)} - e^{j\theta_i}\sqrt{\kappa_i}a_i \tag{2b}$$

Since the input-signal frequency is $f = \omega/2\pi$, the field everywhere oscillates as $e^{-j\omega t}$, and thus $da_i/dt = -j\omega a_i$. The relationship between the incident and output waves of the ith cavity can be expressed as

$$S_{f,out}^{(i)} = \frac{j(\omega_i - \omega) + \gamma_i - \kappa_i}{j(\omega_i - \omega) + \gamma_i}S_{f,in}^{(i)} - \frac{\kappa_i}{j(\omega_i - \omega) + \gamma_i}S_{b,out}^{(i)} \tag{3a}$$

$$S_{b,in}^{(i)} = \frac{\kappa_i}{j(\omega_i - \omega) + \gamma_i}S_{f,in}^{(i)} + \frac{j(\omega_i - \omega) + \gamma_i + \kappa_i}{j(\omega_i - \omega) + \gamma_i}S_{b,out}^{(i)} \tag{3b}$$

When the microwave currents are only inputted from the left port ($S_{b,\,in}^{(i)} = 0$), the transmission and reflection coefficients of a single ISRR coupled with a microstrip-line waveguide are derived as

$$t_i(\omega) = \frac{j(\omega_i - \omega) + \gamma_i}{j(\omega_i - \omega) + \gamma_i + \kappa_i} \tag{4a}$$

$$r_i(\omega) = -\frac{\kappa_i}{j(\omega_i - \omega) + \gamma_i + \kappa_i} \tag{4b}$$

If the side-coupled resonators can be regarded as frequency-dependent loss mirrors $[r_i^2(\omega) + t_i^2(\omega) < 1]$, the propagating waves in the microstrip line should satisfy the relationship in the steady state:

$$S_{b,\,in}^{(i)} = S_{b,\,out}^{(i+1)} e^{j\varphi_i}, \quad S_{f,\,in}^{(i+1)} = S_{f,\,out}^{(i)} e^{j\varphi_i} \tag{5}$$

Hence, the transmission behaviors in each resonator can be expressed as:

$$\begin{pmatrix} S_{b,in}^{(2)} \\ S_{f,out}^{(2)} \end{pmatrix} = V \begin{pmatrix} S_{f,in}^{(1)} \\ S_{b,out}^{(1)} \end{pmatrix} \tag{6a}$$

with

$$V = \begin{bmatrix} -\dfrac{r_2}{t_2} & \dfrac{1}{t_2} \\ 1 + \dfrac{r_2}{t_2} & \dfrac{r_2}{t_2} \end{bmatrix} \begin{bmatrix} 0 & e^{j\varphi} \\ e^{-j\varphi} & 0 \end{bmatrix} \begin{bmatrix} -\dfrac{r_1}{t_1} & \dfrac{1}{t_1} \\ 1 + \dfrac{r_1}{t_1} & \dfrac{r_1}{t_1} \end{bmatrix} \tag{6b}$$

where $\varphi = (\omega/c_{\text{eff}})d$ is the phase difference with $c_{\text{eff}} = c/\sqrt{\epsilon_{\text{eff}}}$, the effective dielectric constant $\epsilon_{\text{eff}}$ [39]. The value of $\epsilon_{\text{eff}} = 4.15$ was extracted from the simulation result. The forward (backward) direction of the incident microwaves is designated as coming from the left (right). Hence, the complex scattering parameters for the forward transmission $S_{21}$, backward transmission $S_{12}$, forward reflection $S_{11}$, and backward reflection $S_{22}$ are finally expressed as:

$$S_{11} = \left.\frac{S_{b,\,out}^{(1)}}{S_{f,\,in}^{(1)}}\right|_{S_{b,in}^{(2)}=0} = \frac{r_1 + r_2 t_1 e^{2i\varphi_1} + r_1 r_2 e^{2i\varphi_1}}{1 - r_1 r_2 e^{2i\varphi_1}} \tag{7a}$$

$$S_{12} = \left.\frac{S_{b,\,out}^{(1)}}{S_{b,\,in}^{(2)}}\right|_{S_{f,in}^{(1)}=0} = \frac{t_1 t_2 e^{i\varphi_1}}{1 - r_1 r_2 e^{2i\varphi_1}} \tag{7b}$$

$$S_{12} = \left.\frac{S_{b,\,out}^{(1)}}{S_{b,\,in}^{(2)}}\right|_{S_{f,in}^{(1)}=0} = \frac{t_1 t_2 e^{i\varphi_1}}{1 - r_1 r_2 e^{2i\varphi_1}} \tag{7c}$$

$$S_{22} = \left.\frac{S_{f,\,out}^{(2)}}{S_{b,\,in}^{(2)}}\right|_{S_{f,in}^{(1)}=0} = \frac{r_2 + r_1 t_2 e^{2i\varphi_1} + r_1 r_2 e^{2i\varphi_1}}{1 - r_1 r_2 e^{2i\varphi_1}} \tag{7d}$$

From Eqs. (7a), (7b), (7c), and (7d), one can obtain the forward (backward) transmissivity, reflectivity, and forward absorptivity, as $T_{f(b)} = |S_{21(12)}|^2$, $R_{f(b)} = |S_{11(22)}|^2$, $A_{f(b)} = 1 - T_{f(b)} - R_{f(b)}$, respectively. From a reflection point of view ($S_{11}$ and $S_{22}$), it is crucial to have balanced decay rates, $\kappa_1 \sim \kappa_2$, in order to achieve a balanced resonance similar to a Fabry–Pérot. This balance is important because the reflection at the first or second ISRR is directly proportional to the corresponding decay rate $\kappa_1$ or $\kappa_2$, as described in Eq. (4b). The presence of this balanced Fabry–Pérot-like resonance enables the observation of a zero reflection condition [40].

For the validation of these analytical equations, we already showed the comparison of the analytical calculation of $S_{21}$ with the CST simulation, as displayed in Fig. 2(b). The fitting of the analytical model to the simulation data were in good agreement in terms of their resonance (peak) positions and powers, with the following fitting parameters : $\gamma_1 = 0.149$ GHz, $\gamma_2 = 0.073$ GHz, $\kappa_1 = 6.45$ GHz, and $\kappa_2 = 5.236$ GHz. The phase $\varphi$ between the dual ISRRs satisfies mπ (m=1,2, …) [33]. As shown in Fig. 2(b), the transparency peak for $d$ = 18 mm is positioned at the central frequencies between the resonance frequencies, 4.03 and 4.04 GHz as obtained from the simulation and analytical model, respectively. According to Eq. (6b), the phase term φ approaches 0.99 π (~mπ) at 4.04 GHz when $d$ = 18 mm.

Liu et,al revealed that detuning the resonator frequency offers a practical approach to control not only the quality factor of the transparency resonance peak but also the amplitude of transmission [33]. In our case, when $d$ = 20 mm, the transparency-resonance peak becomes asymmetric and tilted, which is consistent with the experimental data shown in Fig. 2(a). The observed frequency shift in our dual resonators validates the satisfaction of the Fabry–Pérot-like model [33,36,41–44]. In a resonant structure coupled to a single radiation channel (which corresponds to the microstrip line in our case), it is well-established that unity reflection occurs near the resonance frequency if there are no other losses. This occurs because the direct transmission and resonant radiation interfere and completely cancel each other [45]. In contrast to the previously reported Fabry–Pérot type interaction with identical resonant 'mirrors' ($\omega_1 = \omega_2$, $\varphi \neq 0$), our system represents a more generalized coupling model system with non-identical resonant mirrors and an arbitrary phase $\varphi$. This allows for unification of different types of Fabry–Pérot and Friedrich-Wintgen systems ($\omega_1 \neq \omega_2$, $\varphi = 0$) [46]. A more generalized coupling model using two non-identical magnons coupled to the same microwave transmission line was already reported in [40]. However, unlike that study, which focused on perfect zero reflection with nearly full absorption and nearly full transmission, our system provides an experimental observation of nearly zero reflection and nearly perfect absorption, exhibiting non-reciprocal responses.

Next, we move on to the analytical model calculation of $|S_{11(22)}|$ reflection spectra compared with the $|S_{21(12)}|$ transmission spectra on the plane of inter-distance and frequency,

within an extended range of $f$ = 3.6 ~ 4.4 GHz and $d$ = 9 ~ 30 mm, as shown in Fig. 5(a). This extended range calculation is an advantage of our analytical model study. We also obtained the reflectivities ($R_{11}$ and $R_{22}$), absorptivity ($A_{12}$ and $A_{21}$), and transmittivities ($T_{21}$ and $T_{12}$), as displayed in Fig. 5(b). The wide inter-distance range used in the analytical model calculation facilitates an understanding of the role of traveling waves interacting the dual ISRRs placed at specific inter-distances. The periodic responses of transmission and reflection indicate that the phase of traveling waves plays a crucial role in signal transmission, reflection, and absorption. It is clearly visible that strong non-reciprocity in both reflection and absorption exists in specific inter-distance regions of $d$ = 18 ~ 20 mm. Particularly at $d \approx 20$ mm, we observed a perfect unidirectional reflectionless phenomenon. The forward reflectivity is high, as depicted in the top left side of Fig. 5(b), whereas there is almost zero reflection in the backward direction, as depicted in the bottom left side of Fig. 5(b). This unidirectional reflectionless response with strong asymmetric reflection is a characteristic feature often seen in PT-symmetric structures.

Furthermore, we discovered unidirectional absorption behavior, where the forward direction signal experiences a combination of transmission and reflection between the two ISRRs, without being absorbed. Conversely, the backward direction signal was almost entirely absorbed, as shown in the bottom middle of Fig. 5(b), and this absorption occurs between the two ISRRs without reflection. This asymmetric reflection can be attributed to two aspects: First, the inherent asymmetry in the geometry of our sample affects the different coupling mechanism between the individual resonators and the microstrip line. The different coupling mechanisms indicated that individual resonators have distinct intrinsic and extrinsic decay rates, as discussed in analytical modeling section. Second, the phase interference mediated by traveling waves contributes to the observed reflection asymmetry. The achievement of perfect electromagnetic absorption via light-matter interactions is of great significance as it enables the maximization of energy conversion and information transfer. This characteristic holds substantial practical promise, particularly in the design and implementation of photon detectors/imagers [47–49] and stealth technology [50]. Thus, the observation of unidirectional absorption and complete suppression of reflection provides valuable guidance for effectively regulating microwave propagation and absorption at specific target frequencies.

Next, in Fig. 6, we plotted minimum or maximum values of reflectivity, absorptivity, and transmittivity versus $d$, according to the analytical model, with those of experimental data for $d$ =18, 19, and 20 mm. The transmittivity leads to $T_{21} = T_{12}$ because our photonic system exhibits symmetrical transmission. The reflectivity $R_{11}$ (blue solid line) and $R_{22}$ (red dashed line) and absorptivity $A_{21}$ (blue solid line) and $A_{12}$ (red dashed line) demonstrate a complementary oscillatory tendency versus inter-distance. All experimental data points, depicted by symbols, show good agreement with the analytical model.

Figures 6(b) and 6(c) show that the absorptivity mirrors the reflectivity ones in terms of inverse magnitude, with absorption being weak in one direction and strong in the opposite direction in two regions when the inter-distance approaches $d$= 17 mm and 20 mm, respectively. Furthermore, when the inter-distance satisfying the maximum transparency condition, a specific inter-distance exists, here $d$ = 18.3 mm, at which the transmission (reflection) magnitude for both directions is equal. This particular inter-distance fosters reciprocal behavior in both directions and closely approximates the Fabry–Pérot resonant condition inter-distance ($d$ = 18 mm), leading to enhanced signal transmission. For an inter-distance of approximately 18 mm, it is evident that not only can perfect mirror-like symmetry conditions be achieved, but also there is strong evidence exists for the high magnitude of transparency at $d$ = 18 mm in the experimental results. It is noteworthy point out that, as is well-known for two-port configuration [51], strict perfect absorption cannot be achieved due to perfect zero reflection and simultaneously zero transmission cannot be satisfied. However, in our system, when the unidirectional reflectionless condition is satisfied, almost perfect absorption in one direction was realized for specific inter-distances. This behavior can be attributed to non-Hermitian physics, substantiated by calculating the non-Hermitian scattering matrix S as described by [52]:

$$S = \begin{pmatrix} t & r_b \\ r_f & t \end{pmatrix}$$

where, $t$ denotes the transmission amplitude, while $r_f$ and $r_b$ correspond to the amplitudes of reflection in the forward and backward directions, respectively. The eigenvalues of S are given as $S_\pm = t \pm \sqrt{r_f r_b}$ with the corresponding eigenstates $\psi_\pm = (1, \pm\sqrt{r_f/r_b})$ at $r_b \neq 0$. Due to the non-Hermitian nature of scattering matrix S, it gives rise to a pair of complex eigenvalues and their corresponding eigenvectors, which can coalesce to form exceptional points (EPs). These EPs represent degenerate scattering states characterized by unidirectional, reflectionless propagation in either forward ($r_f = 0$, $r_b \neq 0$) or backward ($r_f \neq 0$, $r_b = 0$) direction.

In Fig. 6(d), we plotted the difference between the upper ($S_+$) and lower ($S_-$) values of the real parts of eigenvalues of the scattering matrix S in a range of $d = 9\sim 30$ mm with a step of 0.1 mm. When the real parts of $S_+$ and $S_-$ coalesce (i.e., the difference is zero), it indicates the emergence of non-hermiticity in our system under specific conditions. However, when the real parts of $S_+$ and $S_-$ are contrasting, it indicates the Hermiticity of our system. Therefore, our system exhibits a similar unidirectional reflectionless behavior. It is worth noting that while non-reciprocal reflection behaviors are present in the range of $d = 17.4 \sim 19.2$, it is only at the inter-distance of $d \approx 17$ mm and $d \approx 20$ mm that we observed nearly zero reflection in one direction and nonzero reflection in the opposite direction (see pink box). Our results indicate that the presence of non-reciprocal reflection properties alone does not guarantee perfect reflectionless and non-reciprocal behavior. Only when the conditions of a non-Hermitian system are satisfied, it exhibits significant unidirectional reflectionless properties. Once more, we emphasize that unidirectional reflectionlessness, together with maximized absorption, is of utmost importance from the application point of view.

**IV. CONCLUSIONS**

We found electromagnetically-induced-transparency (EIT)-like spectra and unidirectional reflectionless absorption in a dual-ISRR system with a single microstrip feeding line. Our analytical model, derived from a temporal coupled mode theory, revealed that the dual-resonator system can be regarded as an array of frequency-filtering lossy mirrors. The transmission spectra of the dual ISRRs exhibit an EIT-like symmetric profile when the retardation phase between the dual ISRRs satisfies the condition of Fabry–Pérot resonance at a specific inter-distance of $d$ =18 mm. Additionally, we demonstrated that the dual resonators exhibiting unidirectional reflectionless phenomena due to traveling wave induced coupling between them can serve as a non-Hermitian system. The unidirectional reflectionless effect was achieved at exceptional points by adjusting the inter-distance according to the analytical model calculation. We observed experimentally considerable asymmetry of the reflectivity, such as $R_{11}$ ($R_{22}$) ~ 0.069 (0.41), 0.63 (0.015), and 0.84 (0.24) at $d$ = 18, 19, and 20 mm, respectively. We also found high asymmetry of the absorptivity, $A_{12}$ ($A_{21}$) ~ 0.75 (0.42), 0.25 (0.87), and 0.14 (0.74) for $d$ =18, 19, and 20 mm, respectively. These measured values indicate non-reciprocal (unidirectional) absorptive behaviors of the forward and backward signal propagations through the dual photonic resonators. Our work not only provides a better understanding of traveling-wave-induced indirect coupling between dual resonators separated largely at specific inter-distances without any coupling with magnetic materials, but also can offer guidelines for practical implications in microwave circuits and quantum information devices.

## ACKNOWLEDGMENTS

SK Hynix Inc. supported this research through the strategic industry-university support program (grant No.0668-20210278). The Institute of Engineering Research at Seoul National University provided additional research facilities for this work.

## DATA AVAILABILITY

Supporting data pertinent to the findings of this study from the corresponding author are available upon reasonable request.

## References

[1] O. J. F. Martin, Molding the Flow of Light with Metasurfaces, 2019 URSI Asia-Pacific Radio Sci. Conf. AP-RASC, New Delhi, India, (2019).

[2] A. A. Maznev, A. G. Every, and O. B. Wright, Wave Motion **50**, 776 (2013).

[3] R. Fleury, D. L. Sounas, and A. Alù, Phys. Rev. Lett. **113**, 023903 (2014).

[4] P. A. Deymier, *Acoustic Metamaterials and Phononic Crystals*, Vol. 173 (Springer-Verlag, Berlin, 2013).

[5] S. Weis, R. Rivière, S. Deléglise, E. Gavartin, O. Arcizet, A. Schliesser, and T. J. Kippenberg, Science 330, 1520 (2010)

[6] Y. Sun, W. Tan, H. Q. Li, J. Li, and H. Chen, Phys. Rev. Lett. **112**, 143903 (2014).

[7] M. Horodynski, M. Kühmayer, C. Ferise, S. Rotter, and M. Davy, Nature **607**, 281 (2022).

[8] P. Tassin, L. Zhang, R. Zhao, A. Jain, T. Koschny, and C. M. Soukoulis, Phys. Rev. Lett. **109**, 187401 (2012).

[9] D. Jalas, A. Petrov, M. Eich, W. Freude, S. Fan, Z. Yu, R. Baets, M. Popović, A. Melloni, J. D. Joannopoulos, and M. Vanwolleghem, Nat. Photonics 7, 579 (2013).

[10] Z. Lin, H. Ramezani, T. Eichelkraut, T. Kottos, H. Cao, and D. N. Christodoulides, Phys. Rev. Lett. **106**, 213901 (2011).

[11] L. Feng, Y. L. Xu, W. S. Fegadolli, M. H. Lu, J. E. B. Oliveira, V. R. Almeida, Y. F. Chen, and A. Scherer, Nat. Mater. **12**, 108 (2013).

[12] L. Feng, Z. J. Wong, R. M. Ma, Y. Wang, and X. Zhang, Science **346**, 972 (2014).

[13] M. Kang, F. Liu, T.-F. Li, Q.-H. Guo, J.Li, and J.Chen, Opt. Lett. 38, (2013)

[14] B. Peng, S. K. Özdemir, F. Lei, F. Monifi, M. Gianfreda, G. L. Long, S. Fan, F. Nori, C. M. Bender, and L. Yang, Nat. Phys. **10**, 394 (2014).

[15] C. Wang, X. Jiang, G. Zhao, M. Zhang, C. W. Hsu, B. Peng, A. D. Stone, L. Jiang, and L. Yang, Nat. Phys. **16**, 334 (2020).

[16] L. Feng, M. Ayache, J. Huang, Y.-L. Xu, M.-H. Lu, Y.-F. Chen, Y. Fainman, and A. Scherer, Science 333, 729 (2011).

[17] X. Zhu, L. Feng, P. Zhang, X. Yin, and X. Zhang, Opt. Lett. **38**, 2821 (2013).

[18] D. L. Sounas, R. Fleury, and A. Alù, Phys. Rev. Appl. **4**, 1 (2015).

[19] R. Bai, C. Zhang, X. Gu, X. R. Jin, Y. Q. Zhang, and Y. Lee, Sci. Rep. **7**, 2 (2017).

[20] R. Bai, C. Zhang, X. Gu, X. R. Jin, Y. Q. Zhang, and Y. Lee, Appl. Phys. Express **10**, (2017).

[21] Y. Fu, Y. Xu, and H. Chen, Opt. Express **24**, 1648 (2016).

[22] M. Sarlsaman, Phys. Rev. A **95**, 013806 (2017).

[23] H. Yin, R. Bai, X. Gu, C. Zhang, G. R. Gu, Y. Q. Zhang, X. R. Jin, and Y. P. Lee, Opt. Commun. **414**, 172 (2018).

[24] S.-H.G. Chang and C.-Y.Sun, Opt. Express 24, (2016)

[25] Z. He, L. Li, W. Cui, Y. Wang, W. Xue, H. Xu, Z. Yi, C. Li, and Z. Li, New J. Phys. **23**, (2021).

[26] B. Bhoi, B. Kim, S. H. Jang, J. Kim, J. Yang, Y. J. Cho, and S. K. Kim, Phys. Rev. B **99**, 134426 (2019).

[27] N. K. Tiwari, A. Sharma, S. P. Singh, M. J. Akhtar, and A. Biswas, Int. J. RF Microw. Comput. Eng. **30**, 1 (2020).

[28] J. D. Baena, J. Bonache, F. Martin, R. M. Sillero, F. Falcone, T. Lopetegi, M. A. G. Laso, J. Garcia-Garcia, I. Gil, M. F. Portillo, and M. Sorolla, IEEE Trans. Microw. Theory Tech. 53, 1451 (2005).

[29] B. Bhoi, S. H. Jang, B. Kim, and S. K. Kim, J. Appl. Phys. **129**, (2021).

[30] N. Liu, M. Hentschel, T. Weiss, A. P. Alivisatos, and H. Giessen, Science **332**, 1407 (2011).

[31] N. Liu, L. Langguth, T. Weiss, J. Kästel, M. Fleischhauer, T. Pfau, and H. Giessen, Nat. Mater. **8**, 758 (2009).

[32] N. Papasimakis, V. A. Fedotov, N. I. Zheludev, and S. L. Prosvirnin, Phys. Rev. Lett. **101**, 253903 (2008).

[33] H. Lu, X. Liu, and D. Mao Phys. Rev. A 85, 053803 (2012).

[34] X. Yang, M. Yu, D.-L. Kwong, and C. W. Wong Phys. Rev. Lett. 102, 173902 (2009).

[35] R. D. Kekatpure, E. S. Barnard, W.Cai, and M.L. Brongersma Phys. Rev. Lett. 104, 243902 (2010).

[36] Z. Ling, Y. Zeng, G. Liu, L. Wang, and Q. Lin, Opt. Express **30**, 21966 (2022).

[37] W. Suh, Z. Wang, and S. Fan, IEEE J. Quantum Electron. **40**, 1511 (2004).

[38] H. A. Haus, Waves and Fields in Optoelectronics (Prentice-Hall, Englewood Cliffs, NJ, 1984)

[39] Y. Li, V. G. Yefremenko, M. Lisovenko, C. Trevillian, T. Polakovic, T. W. Cecil, P. S. Barry, J. Pearson, R. Divan, V. Tyberkevych et al., Phys. Rev. Lett. 128, 047701 (2022).

[40] J. Qian, C. H. Meng, J. W. Rao, Z. J. Rao, Z. An, Y. Gui, and C. M. Hu, Nat. Commun. **14**, 3437 (2023).

[41] Z. Han and S. I. Bozhevolnyi, Opt. Express **19**, 3251 (2011).

[42] B. Wang, T. Wang, X. Li, X. Han, and Y. Zhu, J. Appl. Phys. **117**, (2015).

[43] G. Cao, H. Li, S. Zhan, H. Xu, Z. Liu, Z. He, and Y. Wang, Opt. Express **21**, 9198 (2013).

[44] Q. Xu, S. Sandhu, M. L. Povinelli, J. Shakya, S. Fan, and M. Lipson Phys. Rev. Lett. 96, 123901 (2006).

[45] S. Fan, W. Suh, and J. D. Joannopoulos, J. Opt. Soc. Am. A 20, (2003)

[46] C. W. Hsu, B. Zhen, A. D. Stone, J. D. Joannopoulos, and M. Soljacic, Nat. Rev. Mater. **1**(16048), (2016).

[47] W. Wan, Y. Chong, L. Ge, H. Noh, A. D. Stone, and H. Cao, Science **331**, 889 (2011).

[48] B. Peropadre, G. Romero, G. Johansson, C. M. Wilson, E. Solano, and J. J. García-Ripoll Phys. Rev. A 84, 063834 (2011).

[49] L. Hu and G. Chen, Analysis of Optical Absorption in Silicon Nanowire Solar Cells, ASME Int. Mech. Eng. Congr. Expo. Proc. **8**, 1285 (2007).

[50] K. Iwaszczuk, A. C. Strikwerda, K. Fan, X. Zhang, R. D. Averitt, and P. U. Jepsen, Opt. Express **20**, 635 (2012).

[51] J. W. Rao, P. C. Xu, Y. S. Gui, Y. P. Wang, Y. Yang, Bimu Yao, J. Dietrich, G. E. Bridges, X. L. Fan, D. S. Xue & C.-M. Hu. Nat. Commun. **12**, 1 (2021).

[52] A. Mostafazadeh, Phys. Rev. Lett. **102**, 220402 (2009).

# Figure Legends

FIG. 1. A dual-resonator systems composed of two inverted slit-ring resonators (ISSRs) coupled to a shared microstrip feeding line. The ports #1 and #2 of the feeding line are connected to a vector network analyzer (VNA) for transmitting the input and output signals. The two ISRRs have square-shape with identical dimensions, but different split-gap positions (left) perpendicular to the microstrip line and (right) parallel to it. The ISRRs' dimensions are as follows: a = 5 mm, b = 3.8 mm, and g = 0.4 mm. The ISRRs were fabricated on a high-frequency laminate (CER-10 RF substrate) of the following material parameters: relative dielectric constant, $\varepsilon_r = 10$; dissipation factor, 0.0012 at 10 GHz. The dimensions of the microstrip line were set to be width w = 0.55 mm and length L = 46 mm in order to achieve the characteristic impedance of 50 Ω. Photographs (inset) show the front and back sides of an example real sample.

FIG. 2. (a) Experimentally measured scattering transmission spectra of $|S_{21}|$ and $|S_{12}|$ versus AC current frequency for different inter-distances *d* between the two ISRRs, where *d* = 18,19, and 20 mm along with (b) CST simulation result and numerical calculation of an analytical model. In the contour plot, (c) the model calculation and CST-simulation results were carried out in a range of *d* = 18 ~ 20 mm with a step of 0.2 mm. Green arrows indicate EIT-like peaks.

FIG. 3. (a) Experimentally measured reflection spectra of $|S_{11}|$ and $|S_{22}|$ versus AC current frequency for *d* = 18,19, and 20 mm. (b) Corresponding transmittivity ($T_{21}$, $T_{12}$), absorptivity ($A_{21}$, $A_{12}$), and reflectivity ($R_{11}$, $R_{22}$) (as indicated by blue, red, and black colors), calculated from experimental data of both transmission $|S_{21}|$ and $|S_{12}|$ and reflection $|S_{11}|$ and $|S_{22}|$ spectra according to $T_{21} = |S_{21}|^2$, $T_{12} = |S_{12}|^2$, $R_{11} = |S_{11}|^2$ and $R_{22} = |S_{22}|^2$ and $A_{21} = 1 - R_{11} - T_{21}$ and $A_{12} = 1 - R_{22} - T_{12}$.

FIG. 4. Schematic diagram for an analytical model of the dual ISRRs coupled dissipatively to a microstrip line, where the separation $d$ between the two ISRR resonators can vary. Each symbol used is described in the text.

FIG. 5. (a) Contour plots of analytical calculations of the reflection $|S_{11}|$ and $|S_{22}|$ spectra and the transmission $|S_{12}|$ and $|S_{21}|$ spectra on the $d$-$f$ plane in ranges of $d = 10 \sim 30$ mm and $f = 3.6 \sim 4.4$ GHz. (b) Corresponding reflectivity ($R_{11}$, $R_{22}$), absorptivity ($A_{21}$, $A_{12}$), and transmittivity ($T_{21} \sim T_{12}$). The marked rectangular boxes indicate data in the range of $f = 3.85 \sim 4.0$ GHz and $d = 18 \sim 20$ mm. For the analytical calculation, we used the following parameters: $\gamma_1 = 0.15$ GHz, $\gamma_2 = 0.073$ GHz, $\kappa_1 = 6.45$ GHz, and $\kappa_2 = 5.24$ GHz, along with the corresponding retardation phases.

FIG. 6. Analytical calculations of (a) transmittivity ($T = T_{21} = T_{12}$), (b) reflectivity ($R_{11}$ and $R_{22}$), and (c) absorptivity ($A_{21}$ and $A_{12}$) as a function of inter-distance $d$, compared with experimental data points (symbols) for the cases of $d = 18$, 19, and 20 mm. (d) The difference between the upper ($S_+$) and lower ($S_-$) values of the real parts of eigenvalues of the scattering matrix S, which is calculated for different $d$ values ranging from 9 to 30 mm with a step of 0.1 mm. The vertical pink color ranges corresponding to unidirectional reflectionless absorption. Vertical solid green line corresponding to critical inter-distance ($d$), at which the transmission (reflection) magnitude for both directions is equal.

# Figures

FIG.1

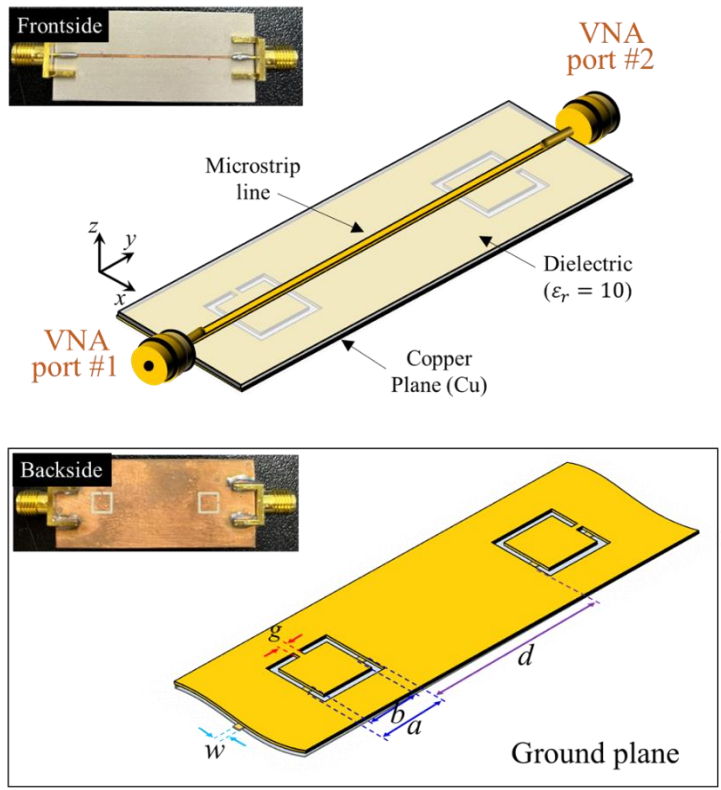

FiG.2

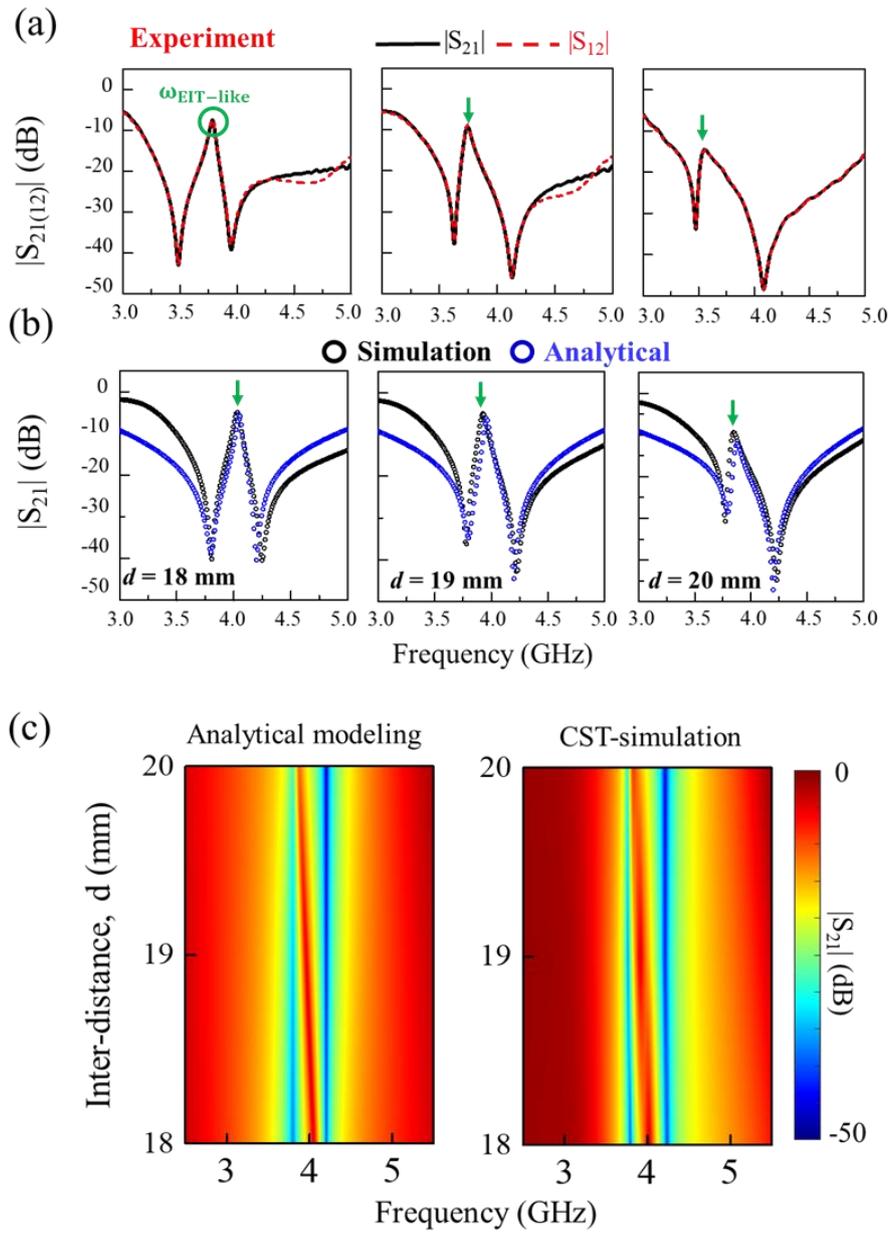

FIG.3

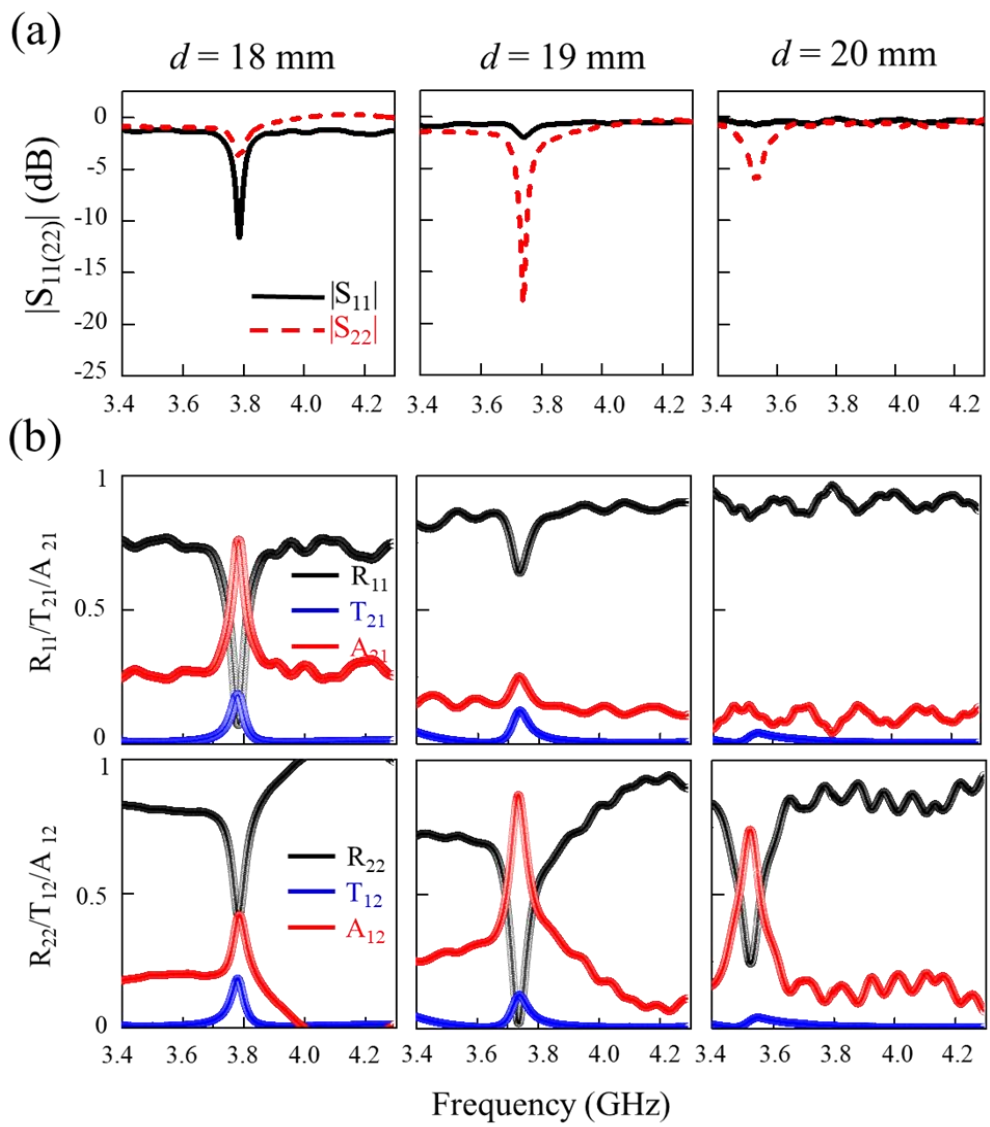

FIG.4

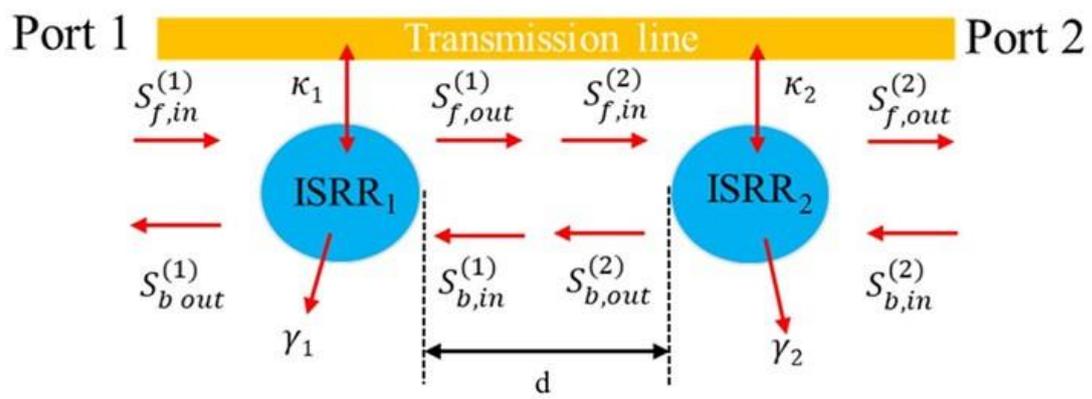

FIG.5

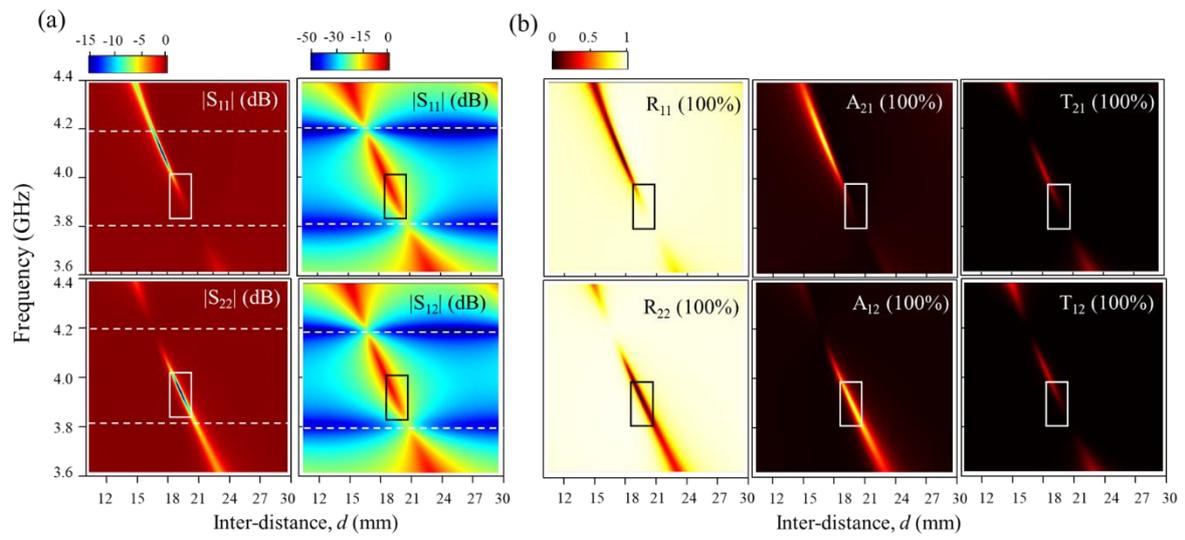

FIG.6

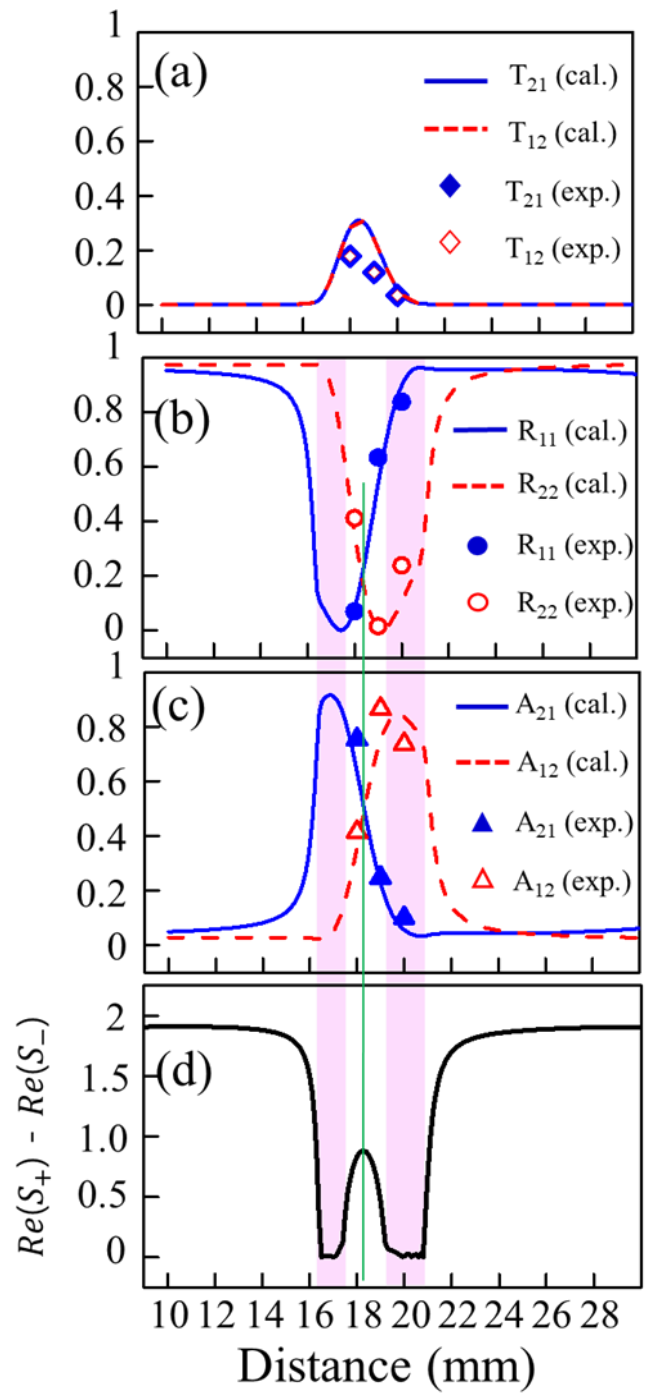